\documentclass[twocolumn,showpacs,aps,pra,floatfix]{revtex4}
\usepackage{epsfig}
\usepackage{amsmath}
\usepackage{amssymb}
\usepackage{bm}
\usepackage{graphicx}
\usepackage{color}

\begin{document}

\title{Electron laser acceleration in vacuum by a quadratically chirped laser pulse}

\author{Yousef I. Salamin}
\affiliation{Department of Physics, American University of Sharjah, POB 26666, Sharjah,
United Arab Emirates}

\author{Najeh M. Jisrawi}
\affiliation{Department of Applied Physics, University of Sharjah, POB 27272, Sharjah, United Arab Emirates}

\date{\today}

\begin{abstract}

Single MeV electrons subjected in vacuum to single high-intensity quadratically-chirped laser pulses are shown to gain multi-GeV energies. The laser pulses are modeled by finite-duration trapezoidal and $\cos^2$ pulse-shapes and the equations of motion are solved numerically. It is found that, typically, the maximum energy gain from interaction with a quadratic chirp is about half of what would be gained from a linear chirp.

\end{abstract}


\maketitle

\section{Introduction}

The technique of chirped pulse amplification (CPA), invented in the mid 1980s \cite{cpa}, has revolutionized laser technology and boosted the laser power outputs by several orders of magnitude in a short time. Current efforts aimed at reaching multi-petawat powers all rely on advances made in CPA \cite{hercules,vulcan,eli}. Laser acceleration of electrons \cite{sal-prl2002,starace,becker,exp2013} and ions \cite{sal-prl2008} to energies that may find application in medicine, the industry or the study of the fundamental forces and the structure of matter in nature, stand to benefit from the strides made in laser technology. See the recent reviews \cite{sal-rev,dipiazza-rev} for more on these extreme laser-matter interactions.

Linearly-chirped pulses have recently been suggested as a means to accelerate particles \cite{sohbatzadeh1,sohbatzadeh2,sohbatzadeh3,li2,sohbatzadeh4,galow,salamin1,li,salamin2}. The main idea stems from the realization that chirping the frequency destroys the symmetry of the pulse sensitively, making energy gain possible by the particle from synchronous interaction with the low-frequency parts of the field \cite{salamin1,li,salamin2}. According to the Lawson-Woodward theorem \cite{lw1,lw2,lw3} particle interaction with a perfectly symmetrical plane-wave pulse results in zero net energy gain. A chirp breaks the symmetry and circumvents the Lawson-Woodward theorem. More on the mechanism of acceleration will be presented shortly. The rest of this Letter presents results of simulations of the dynamics of a single electron submitted to high-intensity laser pulses, chirped linearly and quadratically (the latter being employed here for the first time, to the best of our knowledge). Exit GeV electron kinetic energies, reached as a result of interaction with finite duration pulses, are obtained by solving the relativistic equations of motion numerically.

\begin{figure}[b]
\vskip-0.7cm
\includegraphics[width=8.5cm]{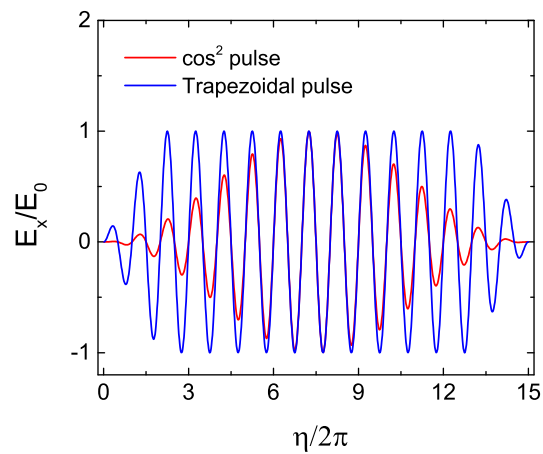}
\caption{(Color online) Normalized electric field of the unchirped plane-wave pulses as functions of the variable $\eta$. The laser wavelength is $\lambda_0=1~\mu$m and $\psi_0 = 0$.}
\label{fig1}
\end{figure}

\section{Theory}

The electron will be treated as a point particle of mass $m$ and charge $-e$, and its relativistic energy and momentum will be denoted, respectively, by ${\cal E}=\gamma mc^2$ and $\bm{p}=\gamma mc\bm{\beta}$, where $\bm{\beta}$ is the velocity of the particle scaled by $c$, the speed of light in vacuum, and $\gamma=(1-\beta^2)^{-1/2}$. The laser fields, on the other hand, will be represented by (SI units)
\begin{equation}\label{EB}
  \bm{E}=\hat{\bm{x}}E_0fg;\quad \text{and} \quad \bm{B}=\hat{\bm{y}}\frac{E_0}{c}fg,
\end{equation}
where $E_0$ is a constant amplitude, $f$ is a suitably-chosen function whose dependence on the space-time coordinates is through the combination $\omega t-kz$, with $\omega$ the frequency and $k=\omega/c$ the wavenumber, and $g$ will denote an appropriate pulse-shape function. Chirping the plane-wave is then tantamount to letting $\omega$ vary with the time in some fashion. In this work, we take $\omega=\omega_0(1+b\eta^n)$, a power chirp, and work with $n=1$ (linear chirp) and $n=2$ (quadratic chirp). Here, $\omega_0$ is the unchirped angular frequency, $b$ is a dimensionless chirp parameter and $\eta=\omega_0t-k_0z$. In terms of the unchirped wavelength, $k_0=2\pi/\lambda_0=\omega_0/c$. The model adopted here introduces the frequency time-variations indirectly, namely, through $t-z/c$ rather than directly via the time $t$. For example, a linear chirp is customarily modeled by $\omega=\omega_0+b_0t$, and so on. Thus our dimensionless chirp parameter may be related to the parameter $b_0$, employed in the general field of signal processing, by $b_0=\omega_0^2b$, with units of $s^{-2}$.

For the plane-wave field function, we will take $f=\sin(\psi_0+\omega t-kz)$. This function takes the following form in the case of a power chirp
\begin{equation}\label{f}
     f = \sin(\psi_0+\eta+b\eta^n),
\end{equation}
where $\psi_0$ is some initial phase which we take equal to zero, for simplicity.

On the other hand, two choices will be made for the pulse-shape. The first choice is a trapezoidal envelope with symmetrical linear turn-on and turn-off, defined by

\begin{equation}\label{g1}
g(\eta) = \left\{
  \begin{array}{ll}
    \eta/5\pi, & \quad \hbox{$0\leq\eta\leq5\pi$;} \\
    1, & \quad \hbox{$5\pi\leq\eta\leq25\pi$;} \\
    -\eta/5\pi+7.5, & \quad \hbox{$25\pi\leq\eta\leq30\pi.$}
  \end{array}
\right.
\end{equation}
As a second choice, we employ the $\cos^2$ envelope
\begin{equation}\label{g2}
    g(\eta) = \cos^2\left[\frac{\pi}{\tau\omega_0}(\eta-\bar{\eta})\right],
\end{equation}
where $\bar{\eta}=15\pi$ has been introduced to make the temporal width $\tau=50$ fs for a wavelength $\lambda_0=1~\mu$m.

\begin{figure}[t]
\includegraphics[width=8.5cm]{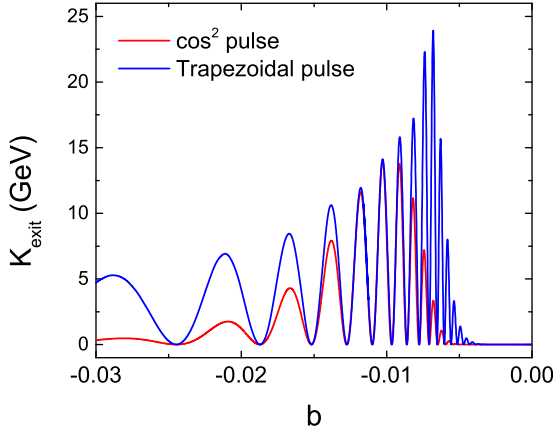}
\caption{(Color online) Exit electron kinetic energy $K_{exit}=mc^2[\gamma_f-1]$, where $\gamma_f=\gamma(\eta_f)$, following interaction with a linear chirp, $\omega=\omega_0(1+b\eta)$, modeled by finite-duration trapezoidal and $\cos^2$ pulses vs the dimensionless chirp parameter $b$. In both cases, $a=3$ ($I\sim1.23\times10^{19}$ W/cm$^2$), $\gamma_0=10$ ($K_0\sim4.6$ MeV), $\lambda_0=1~\mu$m, and $\psi_0 = 0$. Temporal width of the $\cos^2$ pulse is $\tau=50$ fs.}
\label{fig2}
\end{figure}

Plots of the {\it unchirped} normalized electric fields of the laser pulses, modeled by Eqs. (\ref{EB}) and (\ref{f}) and employing (\ref{g1}) and (\ref{g2}) as pulse shapes, are displayed in Fig. \ref{fig1}. Parameters of the two pulse-shapes have been chosen so that the field has the same number of 15 cycles. The fields are almost identical in the central region of the figure. Thus, if they undergo similar chirping (both linear or both quadratic, using the same chirp parameter value) which, in turn, distorts them in the central region, they will present the electron with identical space-time variations and lead to essentially the same energy gain.

\begin{figure}[t]
\includegraphics[width=8.5cm]{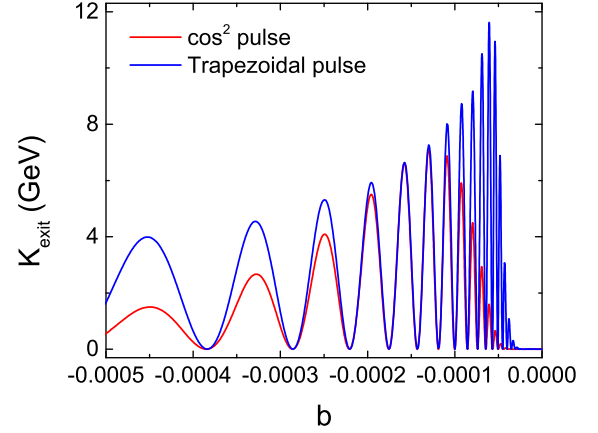}
\caption{(Color online) Same as Fig. \ref{fig2}, but for a quadratic chirp $\omega=\omega_0(1+b\eta^2)$.}
\label{fig3}
\end{figure}

For the sake of solving them numerically, the relativistic Newton-Lorenz equations may be combined to give
\begin{equation}\label{betaeq}
    \frac{d\bm{\beta}}{dt}=\frac{e}{\gamma mc}\left[\bm{\beta}(\bm{\beta}\cdot\bm{E})-(\bm{E}+c\bm{\beta}\times\bm{B})\right],
\end{equation}
Equation (\ref{betaeq}) can be integrated analytically, using $\eta$ as a variable, for only very specialized situations. The main working expressions for the analysis of the particle dynamics have been developed using unchirped \cite{hartemann,salamin3} and chirped pulses \cite{salamin2}. Two constants of the motion were identified \cite{salamin2} which made it possible for some end results to be obtained fully analytically in terms of Fresnel sine and cosine integrals. To save space in this paper, however, we resort to numerical integration instead.

\begin{figure}[t]
\includegraphics[width=8.5cm]{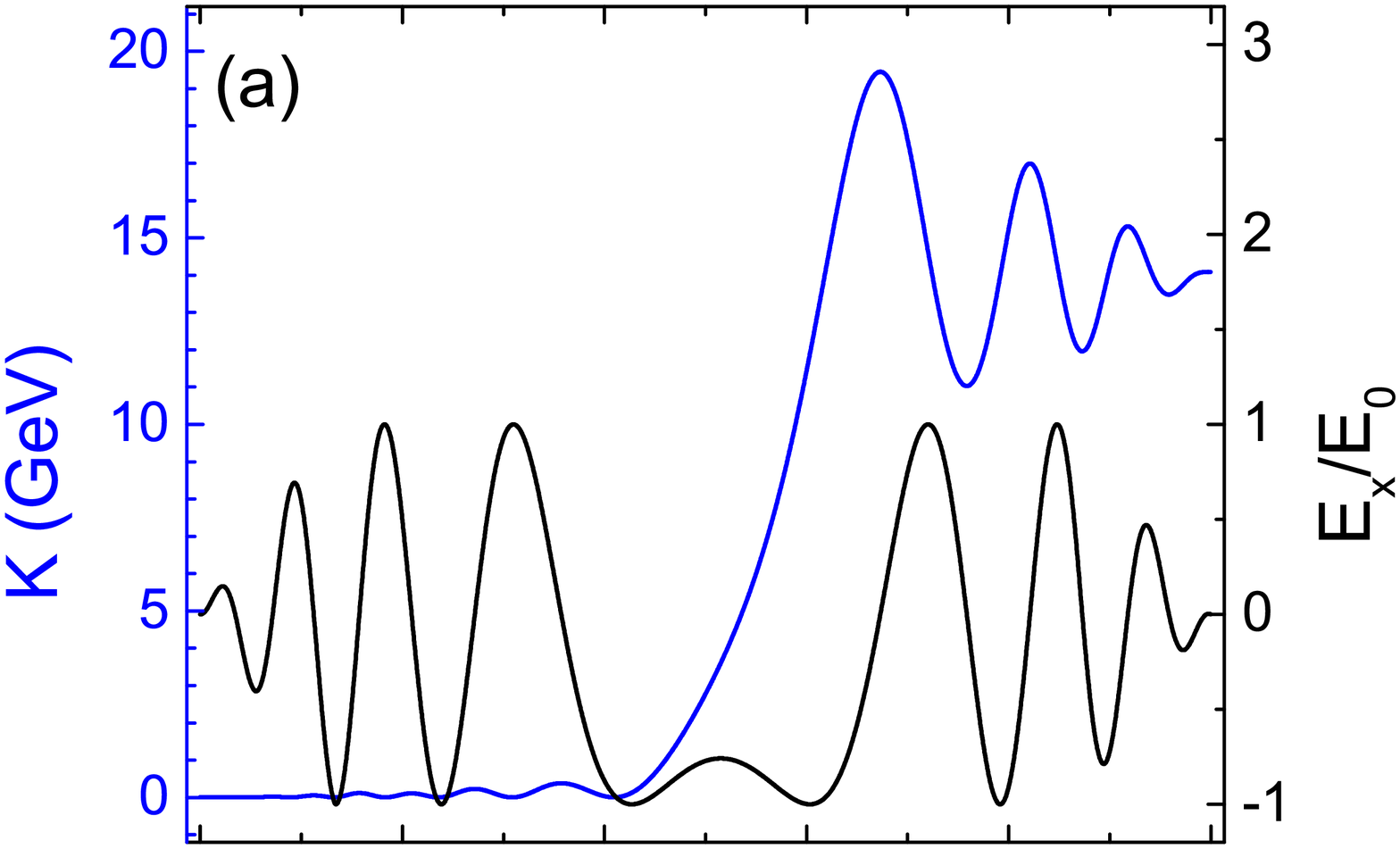}
\vskip-1.5cm
\includegraphics[width=8.5cm]{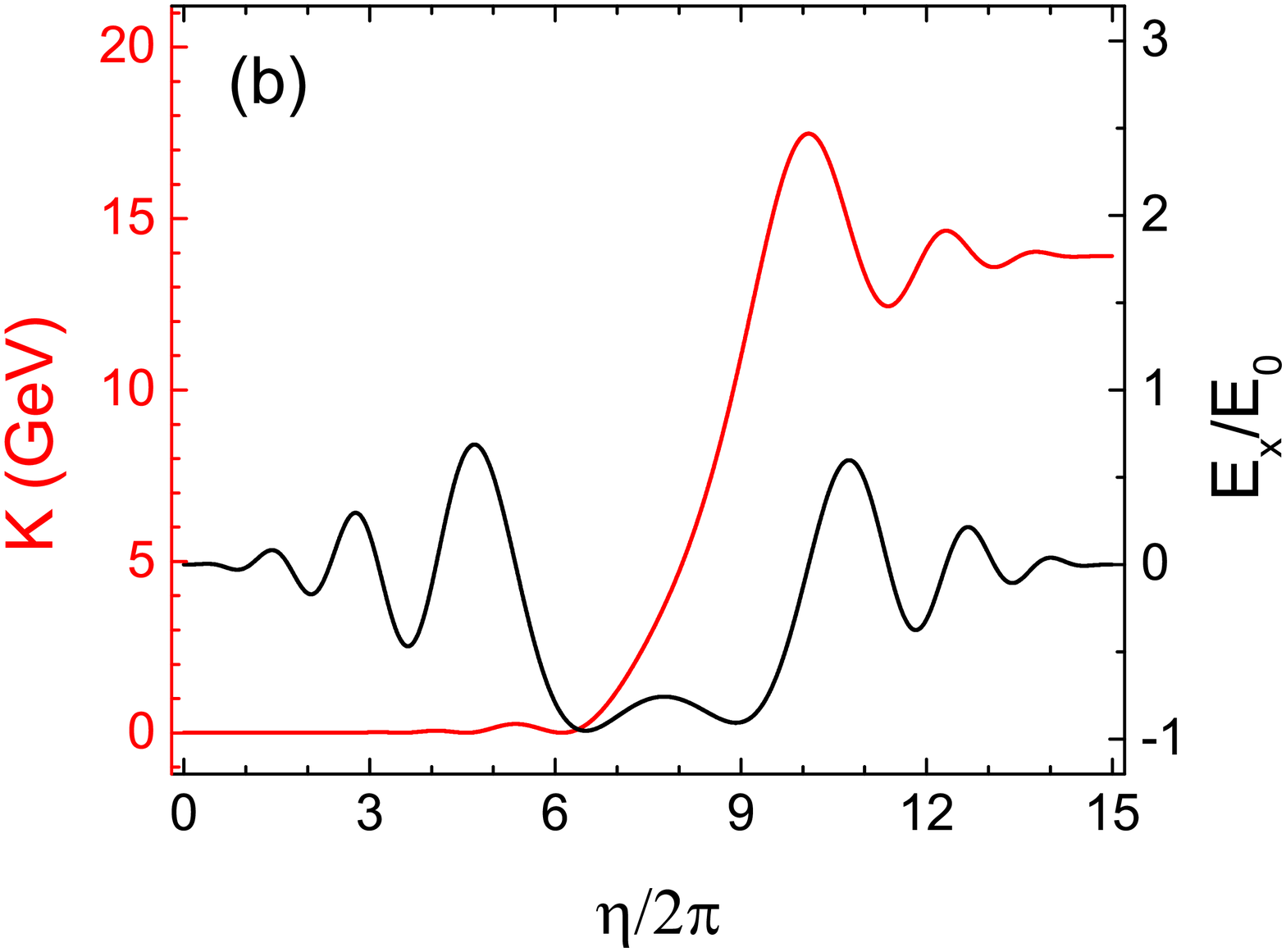}
\caption{(Color online) (a) Evolution in $\eta/2\pi$ of the electron kinetic energy $K(\eta)=mc^2[\gamma(\eta)-1]$ during interaction with a linear chirp, employing $b=-0.0103$ and modeled by a trapezoidal pulse of finite duration. Shown also is the normalized electric field of the linear chirp. All other parameters are the same as in Fig. \ref{fig3}. (b) Same as (a) but for a $\cos^2$ pulse.}
\label{fig4}
\end{figure}

For a single electron injected along the propagation direction of the pulse, the $z-$axis of a cartesian coordinate system in which the $x-$axis serves as the direction of polarization of the laser field, it will be assumed that the front of the pulse catches up with the electron exactly at $t=0$, and precisely at the instant the latter is at the origin of coordinates and moving at a scaled speed $\beta_0$ derived from an injection kinetic energy $K_0=(\gamma_0-1)mc^2$. Thus, one integrates from $\eta_i=0$ (marking onset of the particle-field interaction) to some $\eta_f$ (at which the interaction terminates).

\begin{figure}[t]
\includegraphics[width=8.5cm]{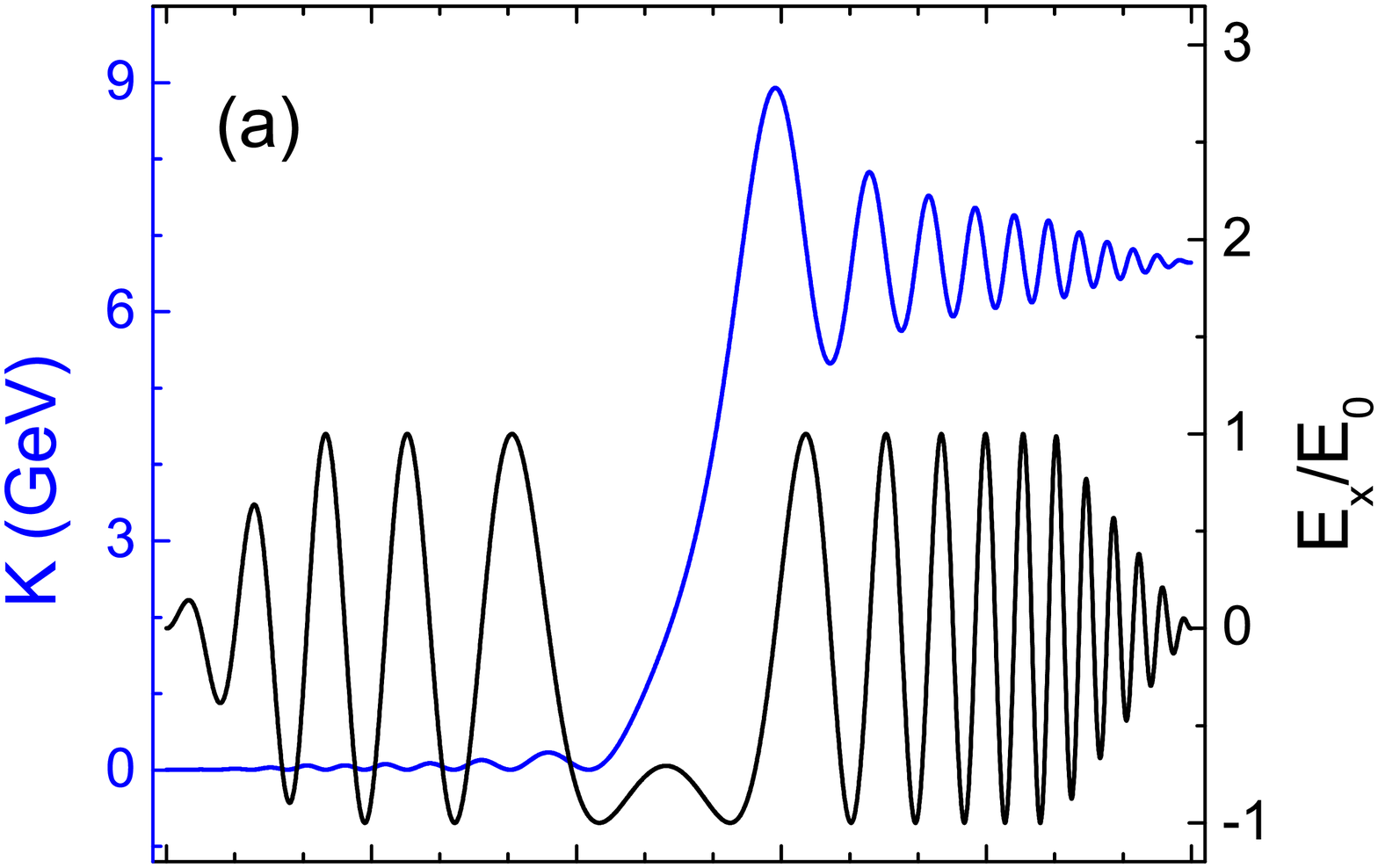}
\vskip-1.5cm
\includegraphics[width=8.5cm]{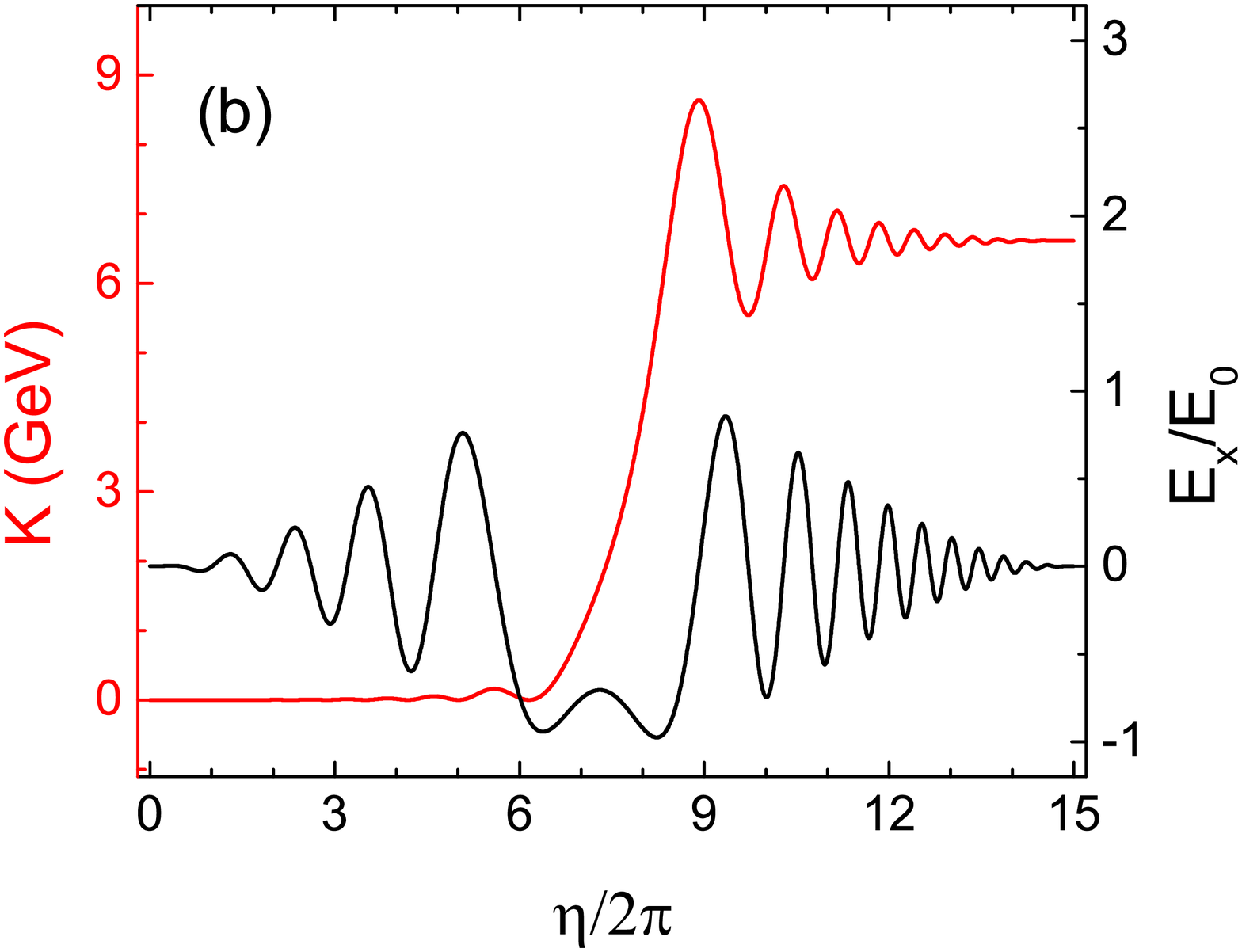}
\caption{(Color online) Same as Fig. \ref{fig4}, but for a quadratic chirp with $b=-0.0001579$. All other parameters are the same as in Fig. \ref{fig4}.}
\label{fig5}
\end{figure}

A single electron injected in the way just described possesses axial relativistic momentum initially. One constant of the motion \cite{salamin2} implies that the $y-$component of the particle's momentum remains zero throughout. Thus the electron follows a 2D trajectory (in the $xz-$plane) \cite{hartemann,salamin3}. Transverse motion is mainly due to $F_x=-eE_x$, while axial motion is governed by $F_z=-e(\bm{v}\times\bm{B})_z$. On the other hand, energy is gained from synchronous interaction with the low-frequency parts of the field brought about by the chirp. The rate of energy gain is $d\varepsilon/dt=-e\bm{v}\cdot\bm{E}$. Note that the low-frequency portions (to be exhibited in Figs. \ref{fig4} and \ref{fig5}, below) lack symmetry. Thus interaction with them results in tremendous energy gain and very little loss.

To achieve energy gains in the GeV regime, relativistic laser field intensities will be required. Although finite-duration pulses will be used in our calculations, yet the dimensionless parameter $a=eE_0/(mc\omega_0)$ familiar from discussions of electron interaction with an infinite plane wave, will be used here. This parameter is related to the intensity $I$ by $a=e\sqrt{2I/c\varepsilon_0}/(mc\omega_0)$, where $\varepsilon_0$ is the permittivity of free space. Inserting values of the universal constants, one gets
\begin{equation}\label{I}
    I\left[\frac{\text{W}}{\text{cm}^2}\right] = 1.36817\times10^{18}\left[\frac{a}{\lambda_0[\text{$\mu$m}]}\right]^2,
\end{equation}
which makes $a^2$ a dimensionless intensity parameter.

\section{Calculations}

Exit electron kinetic energies are presented in Figs. \ref{fig2} and \ref{fig3}, following interaction with the trapezoidal and $\cos^2$ down-chirped ($b<0$) laser pulses, as functions of the chirp parameter $b$. The following discussion applies to both cases of linear chirp (Fig. \ref{fig2}) and quadratic chirp (Fig. \ref{fig3}). In all calculations, the front end of each pulse is assumed to have caught up with the electron at $t=0$ at the coordinate origin. Interaction with the highly symmetrical parts of the pulse leads to no net gain. Substantial net gain results from interaction with those parts of a particular pulse that have been distorted by chirping, as will be illustrated shortly. Thus not all values of $b$ result in gain, while some values do lead to a sharp rise in the electron kinetic energy, which it retains after having been left behind the pulse.

For small ranges of $b$ values, around $b\sim-0.01$ in Fig. \ref{fig2} and $b\sim-0.00015$ in Fig. \ref{fig3}, the energy gain is independent of the pulse-shape employed. This is due to the fact that the central regions of the two pulses are almost identical in the unchirped case and suffer approximately the same distortion after being chirped similarly (same $b$). Since the two pulse-shapes are almost identical, apart from the width of the central region, they lead to kinetic energy minima and maxima at identical values of $b$. Outside these regions of $b$ values, the peaks resulting from interaction with the trapezoidal pulse are substantially higher than the corresponding ones stemming from interaction with the $\cos^2$ pulse. This is due to the difference in pulse width between the two. The trapezoidal pulse is wider and presents the electron with more chance to interact (with a wider distorted, essentially quasi-static, electric field).

While the above discussion has focused on similarities between the linear and quadratic chirp results, there are two important distinctions. The absolute maximum energy gain from a linear chirp is almost twice that from a quadratic chirp. This may be intuitively understood to arise from two factors, both can be inferred from studying Figs. \ref{fig4} and \ref{fig5} below. The quasi-static portion, which we understand to be responsible for the acceleration, is wider in the linear case than in the quadratic. This leads to more particle interaction time with the field and, hence, to more energy gain, in the former compared to the latter. Furthermore, the figures show that the quadratically-chirped fields exhibit more oscillations than the linearly-chirped ones. Recall that interaction with the highly symmetrical oscillations always results in little or no gain (or loss) of energy.

Furthermore, comparable gains are achieved which correspond to values of $|b|$ that are roughly two orders of magnitude smaller in the quadratic case than in the linear case. This could be an advantage the quadratic chirp has over the linear chirp, a speculation on our part that remains to be confirmed experimentally, hopefully sometime soon.

In Figs. \ref{fig4} and \ref{fig5}, evolution of the electron's kinetic energy $K$ with $\eta$ (which implicitly contains dependence upon $t$ and $z$) is shown for values of $b$ picked from Figs. \ref{fig2} and \ref{fig3}, respectively. The value of $b$ chosen in each case corresponds to a point where the calculated gain in energy is the same for both pulse-shapes (pulse-shape independent). With each kinetic energy evolution graph the (chirped) normalized electric field seen by the electron, during interaction with the corresponding pulse, is also shown. Synchronized interaction is clearly evident in all figures. Little gain, if at all, from interaction with the symmetrical field oscillations on the pulse wings, is also clear. More strikingly obvious is the tremendous gain that stems from interaction with the distorted central parts of each pulse. The corresponding kinetic energy evolution curves during interaction with both pulses (same kind of chirp) have the same general structure. While the absolute maximum reached in the case of the trapezoidal pulse is slightly higher, both pulse-shapes result in the same exit energy gain, as has already been explained above.
\\

\section{Concluding remarks}

It has been demonstrated that a single electron submitted to quadratically chirped plane-wave single laser pulses of finite duration, gains multi-GeV energy. For the parameter set used in this paper, it has been shown that the electron exit kinetic energies can reach about one half what would result from interaction with a linear chirp. Of course, with the GeV energies created the question of radiative reactions appears to be relevant \cite{dipiazza2,tamburini}.
Moreover, the values of $b$ employed in our calculations have been quite small, in the range $-0.03<b<0$ (linear) and $-0.0005<b<0$ (quadratic). These ranges correspond roughly to: $-1.06592\times10^{29}~s^{-2}<b_0<0$ (linear) and to $-1.77653\times10^{27}~s^{-2}<b_0<0$ (quadratic). A quick look at Figs. \ref{fig2} and \ref{fig3} may give the impression that a slight deviation from the $b$ value which leads to a typical peak in the exit kinetic energy of the electron would necessarily result in a substantial decrease in the peak energy sought. This casts some doubt about the stability of the laser system used, unless thinking in terms of $b_0$, rather than $b$, is more experimentally relevant. Production of linear and quadratic chirps suitable for experimental realization of the ideas proposed in this work may be currently challenging.

One final remark is in order concerning the quasi-static portion of the pulse field. Strictly speaking, a low-frequency part may contain a zero-frequency component. However, as has recently been demonstrated \cite{galow}, a small band of low frequencies may, in principle, be filtered out without affecting the energy gain appreciably. Moreover, small DC modifications may be added to the wings to allow for the requirement that the integral under the pulse ought to vanish, again without leading to appreciable changes in the results.

\begin{acknowledgments}
YIS acknowledges support for this work from an American University of Sharjah Faculty Research Grant (FRG3).
\end{acknowledgments}


\begin{thebibliography}{99}
 \bibitem{cpa}  D. Strickland and G. Mourou, Opt. Commun. {\bf 56}, 219 (1985).
 \bibitem{hercules} V. Yanovsky, V. Chvykov, G. Kalinchenko, P. Rousseau, T. Planchon, T. Matsuoka, A. Maksimchuk, J. Nees, G. Cheriaux, G. Mourou, and K. Krushelnick, Op. Exp. {\bf 16}, 2109 (2008).

 \bibitem{vulcan} "Central Laser Facility", \url{http://www.clf.rl.ac.uk/}.
 \bibitem{eli} "Extreme Light Infrastructure", \url{http://www.extreme-light-infrastructure.eu/}.
 \bibitem{sal-prl2002} Y. I. Salamin, C. H. Keitel, Phys. Rev. Lett. {\bf 88}, 095005 (2002).
 \bibitem{starace} S. X. Hu and  A. F. Starace, Phys. Rev. Lett. {\bf 88} 245003 (2002).
 \bibitem{becker} Q. Lin, J. Zheng, and W. Becker, Phys. Rev. Lett. {\bf 97}, 253902 (2006).
 \bibitem{exp2013} D. Cline, L. Shao, X. Ding, Y. Ho, Q. Kong, and P. Wang, J. Mod. Phys. {\bf 4}, 1 (2013).
 \bibitem{sal-prl2008} Y. I. Salamin, Z. Harman, C. H. Keitel, Phys. Rev. Lett. {\bf 100}, 155004 (2008).
 \bibitem{sal-rev} Y. I. Salamin, S. X. Hu, K. Z. Hatsagortsyan, and C. H. Keitel, Phys. Rep. {\bf 427}, 41 (2006).
 \bibitem{dipiazza-rev} A. Di Piazza, C. M\"uller, K. Z. Hatsagortsyan, and C. H. Keitel, Rev. Mod. Phys. {bf 84}, 1177 (2012).
 \bibitem{sohbatzadeh1} F. Sohbatzadeh, S. Mirzanejhad, and M. Ghasemi, Phys. Plasma {\bf 13}, 123108 (2006).

 \bibitem{sohbatzadeh2} F. Sohbatzadeh, S. Mirzanejhad, and H. Aku, Phys. Plasma {\bf 16}, 023106 (2009).
 \bibitem{sohbatzadeh3} F. Sohbatzadeh, S. Mirzanejhad, H. Aku, and S. Ashouri, Phys. Plasma {\bf 17}, 083108 (2010).
 \bibitem{li2} J. X. Li, W. P. Zang, and J. G. Tian, Appl. Phys. Lett. {\bf 96}, 031103 (2010).
 \bibitem{sohbatzadeh4} F. Sohbatzadeh, and H. Aku, J. Plasma Phys. {\bf 77}, 39 (2011).

 \bibitem{galow} B. J. Galow, Y. I. Salamin, T. V. Liseykina, Z. Harman, and C. H. Keitel, Phys. Rev. Lett. {\bf 107}, 185002 (2011).
 \bibitem{salamin1} Y. I. Salamin, J. X. Li, B. J. Galow, Z. Harman, and C. H. Keitel, Phys. Rev. A {\bf 85}, 063831 (2012).
 \bibitem{li} J. X. Li, Y. I. Salamin, B. J. Galow, and C. H. Keitel, Phys. Rev. A {\bf 85}, 063832 (2012).
 \bibitem{salamin2} Y. I. Salamin, Phys. Lett. A {\bf 376}, 2442 (2012).
 \bibitem{lw1} J. D. Lawson, IEEE Trans. Nucl. Sci. {\bf 26}, 4217 (1979).
 \bibitem{lw2} P. M. Woodward, J. Inst. Elec. Eng. {\bf 93}, 1554 (1947).
 \bibitem{lw3} R. B. Palmer, in Frontiers of Particle Beams, edited by M.
        Month and S. Turner (Springer-Verlag, New York, 1988) pp. 607--635.

 \bibitem{salamin3} Y. I. Salamin, F. H. M. Faisal, and C. H. Keitel, Phys. Rev. A {\bf 62}, 053809 (2000)

 \bibitem{hartemann} F. V. Hartemann, S. N. Fochs, G. P. Lesage, N. C. Luhmann, J. G. Woodworth,
        M. D. Perry, Y. J. Chen, and A. K. Kerman, Phys. Rev. E {\bf 51}, 4833 (1995).
 \bibitem{dipiazza2} A. Di Piazza, K. Z. Hatsagortsyan, and C. H. Keitel, Phys. Rev. Lett. {bf 102}, 254802 (2009).
 \bibitem{tamburini} M. Tamburini, F. Pegoraro, A. Di Piazza, C. H. Keitel, T. V. Liseykina, and A. Macchi, Nucl. Instr. Meth. Phys. Res. {\bf 653}, 181 (2011).

\end{thebibliography}
\end{document}